\documentstyle[art12,12pt,epsf,epsfig]{article} 
\input epsf

\begin{document}

\def\mathrm{\rm} 
\newcommand{\ra}{\mbox{$\rightarrow$}}
\newcommand{\pr}{\mbox{\small $\prime$}}
\newcommand{\half}{\mbox{$\frac{1}{2}$}}
\newcommand{\third}{\mbox{$\frac{1}{3}$}}
\newcommand{\thovsxt}{\mbox{$\frac{3}{16}$}}
\newcommand{\thoveit}{\mbox{$\frac{3}{8}$}}
\newcommand{\thovfr}{\mbox{$\frac{3}{4}$}}
\newcommand{\eitovth}{\mbox{$\frac{8}{3}$}}
\newcommand{\sxtovth}{\mbox{$\frac{16}{3}$}}
\newcommand{\onovsij}{\mbox{$\frac{1}{\sij}$}}
\newcommand{\onovsi}{\mbox{$\frac{1}{\sigma_i}$}}
                                                   
\newcommand{\tn}{\mbox{$\tau$}}
\def\t{\tn}
\newcommand{\tm}{\mbox{$\tau^-$}}
\newcommand{\tp}{\mbox{$\tau^+$}}
\newcommand{\nut}{\mbox{$\nu_{\tau}$}}
\newcommand{\nutb}{\mbox{$\bar{\nu}_{\tau}$}}
 
\newcommand{\eln}{\mbox{\rm e}}
\newcommand{\elm}{\mbox{\rm e$^-$}}
\newcommand{\elp}{\mbox{\rm e$^+$}}
\newcommand{\nue}{\mbox{$\nu_e$}}
\newcommand{\nueb}{\mbox{$\bar{\nu}_e$}}
 
\newcommand{\mun}{\mbox{$\mu$}}
\newcommand{\mum}{\mbox{$\mu^-$}}
\newcommand{\mup}{\mbox{$\mu^+$}}
\newcommand{\num}{\mbox{$\nu_{\mu}$}}
\newcommand{\numb}{\mbox{$\bar{\nu}_{\mu}$}}
 
\newcommand{\elln}{\mbox{$\ell$}}
\newcommand{\ellm}{\mbox{$\ell^-$}}
\newcommand{\ellp}{\mbox{$\ell^+$}}
\newcommand{\nul}{\mbox{$\nu_{\ell}$}}
\newcommand{\nulb}{\mbox{$\bar{\nu}_{\ell}$}}      
 
\newcommand{\pin}{\mbox{$\pi$}}
\newcommand{\pim}{\mbox{$\pi^-$}}
\newcommand{\pip}{\mbox{$\pi^+$}}
\newcommand{\piz}{\mbox{$\pi^0$}}
\newcommand{\piK}{\mbox{$\pi$}}
\newcommand{\rhon}{\mbox{$\rho$}}
\newcommand{\rhom}{\mbox{$\rho^-$}}
\newcommand{\rhop}{\mbox{$\rho^+$}}
 
\newcommand{\Kstn}{\mbox{\rm K$^{\ast}$}}
\newcommand{\Kstm}{\mbox{\rm K$^{\ast +}$}}
\newcommand{\Kstp}{\mbox{\rm K$^{\ast -}$}}

\newcommand{\ksh}{\mbox{\rm K$^0_s$}}
\newcommand{\Xp}{\mbox{\rm X$^+$}}
\newcommand{\Xm}{\mbox{\rm X$^-$}}

\newcommand{\aonep}{\mbox{\rm a$_1^-$}}
\newcommand{\aonem}{\mbox{\rm a$_1^+$}}
 
\newcommand{\Z}{\mbox{\rm Z$^0$ }}
\def\tt{\mbox{\tp \tm}}
\newcommand{\tautau}{\mbox{\tp \tm}}
\newcommand{\ee}{\mbox{\elp \elm} }
\newcommand{\mumu}{\mbox{\mup \mum} }
\newcommand{\ggmm}{\mbox{$\gamma\gamma$\ra\mumu}}
\newcommand{\led}    {\elln\nulb\nut}
\newcommand{\pid}    {\piK \nut}
 
\newcommand{\tmu}{\mbox{\tn \ra \mun \numb \nut}}
\newcommand{\tel}{\mbox{\tn \ra \eln \nueb \nut}}
\newcommand{\tle}{\mbox{\tn \ra \elln \nulb \nut}}
\newcommand{\tpi}{\mbox{\tn \ra $\pi$ \nut}}
\newcommand{\tpiK}{\mbox{\tn \ra \piK \nut}}
\newcommand{\tro}{\mbox{\tn \ra \rhon\nut}} 
\def\thp{\t \ra h\pi^0\nu}
\newcommand{\tkstr}{\mbox{\tn \ra K$^{\ast}$ \nut}}
\newcommand{\taone}{\mbox{\tn \ra a$_1$\nut}}
\newcommand{\thnpiz}{\mbox{\tn \ra h $>$2\piz \nut}}
\newcommand{\eemm}{\mbox{\ee \ra \mumu} }
\newcommand{\eett}{\mbox{\ee \ra \tautau} }
\newcommand{\pzgg}{\mbox{\piz \ra $\gamma \gamma$}}
\def\p0gg{\pi ^0 \ra \gamma \gamma}   
 
\newcommand{\efswsq}
{\mbox{$\sin^2\theta_{W}^{\mbox{\small \it eff}}$}}
\def\swsq{\sin^2\theta_W} 
\def\cost{\cos\theta}
\newcommand{\cosst}  {\cos^2\theta}
\newcommand{\cst}{\mbox{$\cos\theta$}}
\newcommand{\cstp}{\mbox{$\cos\theta^{\pr}$}}
\newcommand{\csts}{\mbox{$\cos\theta^{\ast}$}}
\newcommand{\cpsi}{\mbox{$\cos\psi$}}
\newcommand{\sspsi}{\mbox{$\sin^2\psi$}}
\newcommand{\cspsi}{\mbox{$\cos^2\psi$}}
\newcommand{\seta}{\mbox{$\sin\eta$}}
\newcommand{\ceta}{\mbox{$\cos\eta$}}
\newcommand{\cseta}{\mbox{$\cos^2\eta$}}
\newcommand{\ctsh}{\mbox{$\cos\frac{\theta^{\ast}}{2}$}}
\newcommand{\stsh}{\mbox{$\sin\frac{\theta^{\ast}}{2}$}}
\newcommand{\cstsh}{\mbox{$\cos^2\frac{\theta^{\ast}}{2}$}}
\newcommand{\sstsh}{\mbox{$\sin^2\frac{\theta^{\ast}}{2}$}}
\newcommand{\act}{\mbox{$|\cos\theta |$}}
\newcommand{\cstsq}{\mbox{$\cos^{2}\theta$}}
\newcommand{\eb}{\mbox{$E_{beam}$}}
\newcommand{\elep}{\mbox{$E_{\ell}$}}
\newcommand{\ehad}{\mbox{$E_h$}}
\newcommand{\ecm}{\mbox{$E_{cm}$}}  
\def\ecma{\langle\ecm\rangle}
\newcommand{\cof}{\mbox{$\chi^2/D.O.F.$}}
\newcommand{\chs}{\mbox{$\chi^2$}}
\newcommand{\xih}{\mbox{$\xi_{had}$}}
\newcommand{\xxi}{\mbox{$x_i$}}
\newcommand{\xxj}{\mbox{$x_j$}}
\newcommand{\xip}{\mbox{$x_i$}}
\newcommand{\xjp}{\mbox{$x_j$}}
\newcommand{\xp}{\mbox{$x$}}
\newcommand{\ptr}{\mbox{$p_{\scriptscriptstyle T}$}}
\newcommand{\xtr}{\mbox{$x_{\scriptscriptstyle T}$}}
\newcommand{\xtp}{\mbox{$x^{\pr}_{\scriptscriptstyle T}$}}
\newcommand{\tpr}{\mbox{$\theta^{\pr}$}}

\newcommand{\fix}{\mbox{$f_i$}}
\newcommand{\fjx}{\mbox{$f_j$}}
\newcommand{\gix}{\mbox{$g_i$}}
\newcommand{\gjx}{\mbox{$g_j$}}
\newcommand{\rxx}{\mbox{${\cal R}_i(\bar{\xxi},\xip)$}}
\newcommand{\rcc}{\mbox{${\cal C}_i(\cos\bar{\theta},\cst)$}}
\newcommand{\mhs}{\mbox{$m_{\pi}^2$}}
\newcommand{\mts}{\mbox{$m_{\tau}^2$}}
\newcommand{\mrs}{\mbox{$m_{\rho}^2$}}
\newcommand{\mpc}{\mbox{$m_{\pi^{\pm}}$}}
\newcommand{\mpz}{\mbox{$m_{\pi^0}$}}
\newcommand{\mt}{\mbox{$m_{\tau}$}}
\newcommand{\mr}{\mbox{$m_{\rho}$}}
\newcommand{\mzsq}{\mbox{$m_{\mbox \rm Z}^2$}}
\newcommand{\mz}{\mbox{$m_Z$}}
\def\mzs{s=m_Z^2}   
\newcommand{\lamt}{\mbox{${\cal A}_{\tau}$}}
\newcommand{\lame}{\mbox{${\cal A}_e$}}
\newcommand{\laml}{\mbox{${\cal A}_{\ell}$}}
\newcommand{\vovat}{\mbox{\rm g$_v^{\tau}$/g$_a^{\tau}$}}
\newcommand{\vovae}{\mbox{\rm g$_v^e$/g$_a^e$}}
\newcommand{\voval}{\mbox{ \rm g$_v^{\ell}$/g$_a^{\ell}$} }
\newcommand{\epstij}
{\mbox{${\cal E}^{\pm}_{ij}(\xip,\xjp,\cst)$}}
\newcommand{\epsti}
{\mbox{${\cal E}^{\pm}_{i}(\xip,\cst)$}}
\newcommand{\epsdi}{\mbox{$\epsilon_{i}(\xip,\cst)$}}
\newcommand{\epsdj}{\mbox{$\epsilon_{j}(\xjp,\cst)$}}
\def\epstpi{{\cal E}^{\pr\pm}_{i}(\xip,\cst)}
\newcommand{\hi}{\mbox{$h^{\pm}_i(\xip)$}}
\newcommand{\hj}{\mbox{$h^{\pm}_j(\xjp)$}}
\newcommand{\beti}{\mbox{$\beta^{\pm}_{i} (\xip,\cst)$}}
\newcommand{\betj}{\mbox{$\beta^{\pm}_{j} (\xjp,\cst)$}}
\def\bbetij{\bar{\beta}^\pm_{ij}}
\def\bbeti{\bar{\beta}^\pm_{i}}
\newcommand{\bntij}{\mbox{$\beta^{non-\tau}_{ij}(\xip,\xjp,\cst)$}}
\newcommand{\bnti}{\mbox{$\beta^{non-\tau}_{i}(\xip,\cst)$}}
\newcommand{\ri}{\mbox{$r^{\pm}_i(\bar{\xxi})$}}
\newcommand{\ric}{\mbox{$\hat{r}^{\pm}_i(\cos\bar{\theta})$}}
 
\newcommand{\ptm}{\mbox{$\langle P_{\tau^-}\rangle$}}
\newcommand{\ptp}{\mbox{$\langle P_{\tau^+}\rangle$}}
\newcommand{\pta}{\mbox{$\langle P_{\tau}\rangle$}}
\newcommand{\pfa}{\mbox{$\langle P_{f}\rangle$}}
\newcommand{\ptat}{\mbox{$\langle P_{\tau}\rangle_{\theta}$}}
\newcommand{\dpta}{\mbox{$\Delta \langle P_{\tau}\rangle$}}
\newcommand{\Pt}{\mbox{\rm P$_{\tau}$}}
\newcommand{\At}{\mbox{\rm A$_{\tau}$}}
\newcommand{\Al}{\mbox{\rm A$_{\ell}$}}
\newcommand{\hnu}{\mbox{\rm h$_{\nu_{\tau}}$}}
\newcommand{\hnup}{\mbox{\rm h$^{\pi}_{\nu_{\tau}}$}}
\newcommand{\hnur}{\mbox{\rm h$^{\rho}_{\nu_{\tau}}$}}
\newcommand{\Ae}{\mbox{\rm A$_e$}}
\newcommand{\Pe}{\mbox{\it P$_e$}}

\newcommand{\ptau}   {P_{\tau}}
\def\pt{P_{\t}}   
\newcommand{\afb}{\mbox{\rm A$_{FB}$}}
\newcommand{\aplfb}{\mbox{\rm A$_{pol}^{FB}$}}
\newcommand{\daplfb}{\mbox{$\Delta$ \rm A$_{pol}^{FB}$}}
\newcommand{\apolf}  {\langle\ptau\rangle^F}
\newcommand{\apolb}  {\langle\ptau\rangle^B}
\newcommand{\sigp}{\mbox{$\sigma_+$}}
\newcommand{\sigm}{\mbox{$\sigma_-$}}
\newcommand{\sigr}{\mbox{$\sigma_R$}}
\newcommand{\sigl}{\mbox{$\sigma_L$}}
\newcommand{\sigtot}{\mbox{$\sigma_{tot}$}}
\newcommand{\sigtotp}{\mbox{$\sigma_{tot}^{\pr}$}}
\newcommand{\spr}{\mbox{$\sigma$}}
\newcommand{\sij}{\mbox{$\sigma_{ij}$}}
\newcommand{\sijp}{\mbox{$\spr_{ij}$}}
\newcommand{\sip}{\mbox{$\spr_i$}}
\def\sigpp{\sigma(++)}
\def\sigpm{\sigma(+-)}
\def\sigmp{\sigma(-+)}
\def\sigmm{\sigma(--)}      
\newcommand{\sigf}   {\sigma^F}
\newcommand{\sigb}   {\sigma^B}  
\newcommand{\ptrans}{\mbox{\bf p$_{\scriptscriptstyle T}$}}
\def\dsigtw{\frac{d^2\sigma_i}{d\cst\:d\xxi}}
\def\dsigtwp{\frac{d^2\sip}{d\cst\:d\xip}}
\def\dsigth{\frac{d^3\sij}{d\cos\theta\:d\xxi\:d\xxj}}
\def\dsigthp{\frac{d^3\sijp}{d\cst\:d\xip\:d\xjp}}

\def\fz{F_0(s)}
\def\fo{F_1(s)}
\def\ft{F_2(s)}
\def\fth{F_3(s)}
\newcommand{\fos}    {F_0(s)}
\newcommand{\fis}    {F_1(s)}
\newcommand{\fts}    {F_2(s)}
\newcommand{\frs}    {F_3(s)}

\def\qe{q_{e}}
\def\qt{q_{\tau}}
\def\ae{a_{e}}
\def\at{a_{\tau}}
\def\ve{v_{e}}
\def\vt{v_{\tau}}
\def\gv{v_{\ell}}
\def\ga{a_{\ell}}
 
\newcommand{\dzero}{\mbox{$| d_0|$}}
\newcommand{\zzero}{\mbox{$| z_0|$}}
\def\d0{|d_0|}
\def\z0{|z_0|}  
\newcommand{\Excess}{\mbox{\rm E$_{excess}$}}
\newcommand{\dphimax}{\mbox{$\delta \phi_{max}$}}
\newcommand{\Nhl}{\mbox{\rm N$^{HC}_{hits/layer}$}}
\newcommand{\Nhc}{\mbox{\rm N$^{HC}_{layers}$}}
\newcommand{\Nhm}{\mbox{\rm N$^{HC/MU}_{layers}$}}
 
\newcommand{\degree} {$^\circ$}
\newcommand{\cth}    {$|\cos\theta|$}
\newcommand{\id}{identification}
\newcommand{\goro}[1]{\multicolumn{2}{c|}{#1}}
\newcommand{\bb}{\mathrm{b}\overline{\mathrm{b}}}
\def\dedx{\mbox{$\mathrm{d}E/\mathrm{d}x$}}
 
\newcommand{\etal}{\mbox{\it et al.,}~}
\newcommand {\beq} {\begin{equation}}
\newcommand {\eeq} {\end{equation}}
\newcommand {\bea} {\begin{eqnarray}}
\newcommand {\eea} {\end{eqnarray}}
\newcommand{\ind}[1]{#1\index{#1}}

\newcommand{\Ntpair}{4328}


\newcommand{\HNU}{-0.93}
\newcommand{\HNUST}{0.10}
\newcommand{\HNUSY}{0.04}


\newcommand{\PIHNU}{-0.81}
\newcommand{\PIHNUST}{0.17}
\newcommand{\PIHNUSY}{0.02}

\newcommand{\ROHNU}{-0.99}
\newcommand{\ROHNUST}{0.12}
\newcommand{\ROHNUSY}{0.04}

\newcommand{\ETA}{-0.13}
\newcommand{\ETAST}{0.47}
\newcommand{\ETASYS}{0.15}

\newcommand{\RHO}{0.69}
\newcommand{\RHOST}{0.13}
\newcommand{\RHOSYS}{0.05}

\newcommand{\XI}{1.02}
\newcommand{\XIST}{0.36}
\newcommand{\XISYS}{0.05}

\newcommand{\XIDELTA}{0.87}
\newcommand{\XIDELTAST}{0.27}
\newcommand{\XIDELTASY}{0.04}


\newcommand{\NURHO}{0.72}
\newcommand{\NURHOST}{0.09}
\newcommand{\NURHOSYS}{0.03}

\newcommand{\NUXI}{1.05}
\newcommand{\NUXIST}{0.35}
\newcommand{\NUXISYS}{0.04}

\newcommand{\NUDX}{0.88}
\newcommand{\NUDXST}{0.27}
\newcommand{\NUDXSYS}{0.04}


\newcommand{\ERHO}{0.71}
\newcommand{\ERHOST}{0.14}
\newcommand{\ERHOSYS}{0.05}

\newcommand{\EXI}{1.16}
\newcommand{\EXIST}{0.52}
\newcommand{\EXISY}{0.06}

\newcommand{\EXIDELTA}{0.85}
\newcommand{\EXIDELTAST}{0.43}
\newcommand{\EXIDELTASY}{0.08}

\newcommand{\META}{-0.59}
\newcommand{\METAST}{0.82}
\newcommand{\METASYS}{0.45}

\newcommand{\MRHO}{0.54}
\newcommand{\MRHOST}{0.28}
\newcommand{\MRHOSYS}{0.14}

\newcommand{\MXI}{0.75}
\newcommand{\MXIST}{0.50}
\newcommand{\MXISY}{0.14}

\newcommand{\MXIDELTA}{0.82}
\newcommand{\MXIDELTAST}{0.32}
\newcommand{\MXIDELTASY}{0.07}

\newcommand{\EHNU}{-0.91}
\newcommand{\EHNUST}{0.31}
\newcommand{\EHNUSY}{0.10}

\newcommand{\MHNU}{-1.02}
\newcommand{\MHNUST}{0.28}
\newcommand{\MHNUSY}{0.05}

\newcommand{\LHNU}{-0.96}
\newcommand{\LHNUST}{0.19}
\newcommand{\LHNUSY}{0.03}

\newcommand{\LHHNU}{-0.94}
\newcommand{\LHHNUST}{0.09}
\newcommand{\LHHNUSY}{0.03}


%
%

\begin{titlepage}
\bigskip 
\begin{flushright}
SLAC-PUB-7333\\
hep-ex/9701020\\
January 28, 1997\\
\end{flushright}
\begin{center}
\bigskip 
 {\Large{\bf
    Measurement of the {\LARGE {\em $\tau$}} Neutrino Helicity \\
    and Michel Parameters in Polarized $\ee$ Collisions}}
\bigskip 
\end{center}
\begin{center}
{\bf The SLD Collaboration$^*$}\\
{\it Stanford Linear Accelerator Center}\\
{\sl Stanford University, Stanford, CA 94309}

\end{center}

\bigskip 
\begin{abstract}
\noindent
 
We present a new measurement of the $\tau$ neutrino helicity $\hnu$ and the
$\tau$ Michel parameters  $\rho$, $\eta$, $\xi$ and the product $ \delta \xi$.
The analysis exploits the highly 
polarized SLC electron beam to extract these  quantities directly from 
a measurement of the $\tn$ decay spectra, using 
the 1993-1995 SLD data sample of $\Ntpair$ 
$e^+e^-\rightarrow Z^0 \rightarrow \tt$ events.
From the decays $\tpiK$ and $\tro$ we obtain
a combined value $\hnu=\HNU \pm \HNUST \pm \HNUSY$.
The leptonic decay channels yield
combined values of
$ \rho= \NURHO \pm \NURHOST \pm \NURHOSYS $,
$\xi=\NUXI \pm \NUXIST \pm \NUXISYS $ and
$\delta\xi= \NUDX \pm \NUDXST \pm \NUDXSYS$.
\end{abstract}


\vspace{1cm}
\begin{center}
 to be published in {\sl Physical Review Letters} {\bf 78}, Number 23, June 1997. 
\end{center} 
\vfill
\end{titlepage}

\clearpage
\small
\begin{center}                          
%
%
%
  \def\iADEL{$^{(1)}$}
  \def\iBOL{$^{(2)}$}
  \def\iBU{$^{(3)}$}
  \def\iBRUN{$^{(4)}$}
  \def\iUCSB{$^{(5)}$}
  \def\iUCSC{$^{(6)}$}
  \def\iCIN{$^{(7)}$}
  \def\iCSU{$^{(8)}$}
  \def\iCOLO{$^{(9)}$}
  \def\iCOL{$^{(10)}$}
  \def\iFER{$^{(11)}$}
  \def\iFRA{$^{(12)}$}
  \def\iILL{$^{(13)}$}
  \def\iLBL{$^{(14)}$}
  \def\iMIT{$^{(15)}$}
  \def\iMASS{$^{(16)}$}
  \def\iMISS{$^{(17)}$}
  \def\iMOSC{$^{(18)}$}
  \def\iNAG{$^{(19)}$}
  \def\iOREG{$^{(20)}$}
  \def\iPAD{$^{(21)}$}
  \def\iPERU{$^{(22)}$}
  \def\iPISA{$^{(23)}$}
  \def\iRUT{$^{(24)}$}
  \def\iRAL{$^{(25)}$}
  \def\iSOGANG{$^{(26)}$}
  \def\iSOONG{$^{(27)}$}
  \def\iSLAC{$^{(28)}$}
  \def\iTENN{$^{(29)}$}
  \def\iTOH{$^{(30)}$}
  \def\iVAND{$^{(31)}$}
  \def\iWASH{$^{(32)}$}
  \def\iWISC{$^{(33)}$}
  \def\iYALE{$^{(34)}$}
  \def\dead{$^{\dag}$}
  \def\andgen{$^{(a)}$}
  \def\andper{$^{(b)}$}
%
%
$^*$
\mbox{K. Abe                 \unskip,\iNAG}
\mbox{K. Abe                 \unskip,\iTOH}
\mbox{T. Akagi               \unskip,\iSLAC}
\mbox{N.J. Allen             \unskip,\iBRUN}
\mbox{W.W. Ash               \unskip,\iSLAC$^\dagger$}
\mbox{D. Aston               \unskip,\iSLAC}
\mbox{K.G. Baird             \unskip,\iRUT}
\mbox{C. Baltay              \unskip,\iYALE}
\mbox{H.R. Band              \unskip,\iWISC}
\mbox{M.B. Barakat           \unskip,\iYALE
% ...
}
\it
  \vskip \baselineskip                   
  \centerline{(\bf The SLD Collaboration)}   
  \vskip \baselineskip                   
%
%
%
  \iADEL
     Adelphi University,
     Garden City, New York 11530 \break
  \iBOL
     INFN Sezione di Bologna,
     I-40126 Bologna, Italy \break
  \iBU
     Boston University,
     Boston, Massachusetts 02215 \break
  \iBRUN
     Brunel University,
     Uxbridge, Middlesex UB8 3PH, United Kingdom \break
  \iUCSB
     University of California at Santa Barbara,
     Santa Barbara, California 93106 \break
  \iUCSC
     University of California at Santa Cruz,
     Santa Cruz, California 95064 \break
  \iCIN
     University of Cincinnati,
     Cincinnati, Ohio 45221 \break
  \iCSU
     Colorado State University,
     Fort Collins, Colorado 80523 \break
  \iCOLO
     University of Colorado,
     Boulder, Colorado 80309 \break
  \iCOL
     Columbia University,
     New York, New York 10027 \break
  \iFER
     INFN Sezione di Ferrara and Universit\`a di Ferrara,
     I-44100 Ferrara, Italy \break
  \iFRA
     INFN  Lab. Nazionali di Frascati,
     I-00044 Frascati, Italy \break
  \iILL
     University of Illinois,
     Urbana, Illinois 61801 \break
  \iLBL
     Lawrence Berkeley Laboratory, University of California,
     Berkeley, California 94720 \break
  \iMIT
     Massachusetts Institute of Technology,
     Cambridge, Massachusetts 02139 \break
  \iMASS
     University of Massachusetts,
     Amherst, Massachusetts 01003 \break
  \iMISS
     University of Mississippi,
     University, Mississippi  38677 \break
  \iMOSC
    Moscow State University,
    Institute of Nuclear Physics
    119899 Moscow, Russia    \break
  \iNAG
     Nagoya University,
     Chikusa-ku, Nagoya 464 Japan  \break
  \iOREG
     University of Oregon,
     Eugene, Oregon 97403 \break
  \iPAD
     INFN Sezione di Padova and Universit\`a di Padova,
     I-35100 Padova, Italy \break
  \iPERU
     INFN Sezione di Perugia and Universit\`a di Perugia,
     I-06100 Perugia, Italy \break
  \iPISA
     INFN Sezione di Pisa and Universit\`a di Pisa,
     I-56100 Pisa, Italy \break
  \iRUT
     Rutgers University,
     Piscataway, New Jersey 08855 \break
  \iRAL
     Rutherford Appleton Laboratory,
     Chilton, Didcot, Oxon OX11 0QX United Kingdom \break
  \iSOGANG
     Sogang University,
     Seoul, Korea \break
  \iSOONG
     Soongsil University,
     Seoul, Korea  156-743 \break
  \iSLAC
     Stanford Linear Accelerator Center, Stanford University,
     Stanford, California 94309 \break
  \iTENN
     University of Tennessee,
     Knoxville, Tennessee 37996 \break
  \iTOH
     Tohoku University,
     Sendai 980 Japan \break
  \iVAND
     Vanderbilt University,
     Nashville, Tennessee 37235 \break
  \iWASH
     University of Washington,
     Seattle, Washington 98195 \break
  \iWISC
     University of Wisconsin,
     Madison, Wisconsin 53706 \break
  \iYALE
     Yale University,
     New Haven, Connecticut 06511 \break
  \dead
     Deceased \break
  \andgen
     Also at the Universit\`a di Genova \break
  \andper
     Also at the Universit\`a di Perugia \break
\rm
\end{center}

We present a study of the fundamental structure of  the charged weak current 
by investigating the energy spectra of 
decay products of polarized $\tau$ leptons at the $\Z$ resonance.
We measure the  $\tau$ 
neutrino helicity and decay Michel parameters, which are 
related~\cite{Fetscher,PDG} to the couplings in $W \to \tau \nut$.
The $\tau$ decay spectra are determined by the nature of the 
decay~\cite{Tsai} and the spin polarization of the taus.
Tau pairs are produced with high longitudinal polarization in collisions of 
highly polarized electrons with positrons at the SLC.

In $\ee \to \tt$, final state $\tau$ polarization results both from
incident beam polarization and from parity violation of the $\Z$ couplings.
Neglecting $\gamma$ exchange and $\gamma-\Z$ interference we 
can write the $\tau$ polarization as:
\beq
  P_\tau(\cst,\Pe) \equiv
\frac{\frac{d\sigma_R}{d\cos\theta}-\frac{d\sigma_L}
{d \cos\theta} }
{\frac{d\sigma_R}{ d\cos\theta}
+ \frac{d\sigma_L}{d\cos\theta}}=
-\frac{ \At + 2 \frac{{\rm A}_e - {\it P}_e}{ 1 -{\rm A}{\it P}_e } 
\frac{\cos\theta}
{ 1 + \cos^2\theta}}
{1 + 2 \At \frac{{\rm A}_e  - {\it P}_e}{ 1 -{\rm A}{\it P}_e }
 \frac{ \cos\theta}{ 1 + \cos^2\theta}},
\eeq
where $\theta$ is the polar angle of the $\tm$ ($\tp$) with respect to the
incident $\elm$ ($\elp$) direction, $\Pe$ is the $\elm$ beam 
polarization,
and $\Ae$ and $\At$~\cite{Al} are the
$\elm$ and $\tau$ parity-violation asymmetry parameters.
This expression is plotted in Fig.~1 for $\Pe=0$ and $\pm 0.8$.
Experiments without beam polarization have relied on spin correlations
between the final state taus [5-8] for measuring the magnitude of the 
parameters 
described above, or the hadronic structure function to measure both the
magnitude and the sign of $\hnu$~\cite{ARGUSHADS,OPALHADS}. 
\begin{figure}[htb]
\centerline{\epsfig{file=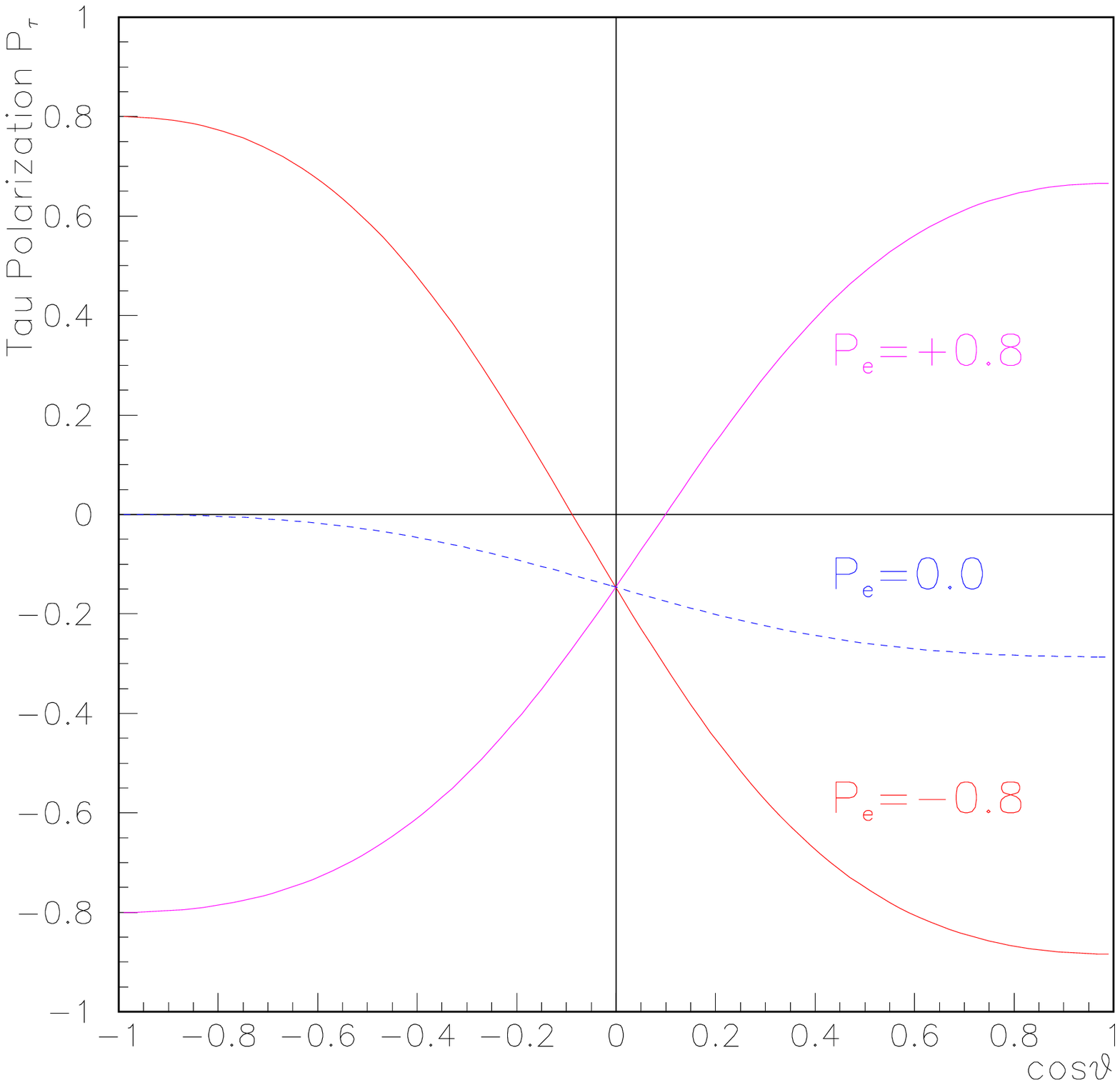
,height=6in,width=6in}}
\caption{$\tau$ polarization vs production angle
 with and without $e^-$ beam polarization.}
\label{jimsplot}
\end{figure}
At the SLC, $\Pt$ is largely determined by the beam polarization and the
production angle, as seen in Fig.~1.
Therefore, this  allows a direct measurement using all individual identified
$\tau$ decays and, unlike  the correlation methods,
provides  the sign as well as the magnitude of  all the polarization-dependent 
parameters.
This is the first such
measurement to be performed with polarized beams.

We study the energy spectra of  pions, rho mesons, electrons 
and muons from $\tau$ decays. 
In the case of a two-body decay, such as $\tpiK$ or $\tro$, the
decay spectrum in the $\tau$ rest frame is mono-energetic, 
so the boosted energy  of the $\pi$ or $\rho$-meson
directly reflects the rest frame decay angle. 
In the case of a three-body decay, such as $\tmu$ or $\tel$, 
the boosted lepton energy reflects  both the 
decay angle and energy in the rest frame. In all cases, we can express
the laboratory energy spectrum in two parts, a polarization-independent
part and a  part that changes sign depending on the 
handedness of the $\tau$: 
$d\Gamma(z,P_{\tau}) \propto (f(z) + \zeta P_{\tau} g(z))dz$.
High $\Pe$ provides enhanced sensitivity to the 
polarization-dependent parameters.

For the $\tpiK$ mode, $z$ is the pion energy scaled by the $\tn$ energy
($z=\frac{E_{\pi}}{E_{\tau}}$), and $f(z)$ and $g(z)$ are given by: 
\beq
f_{\pi}(z)  =  1,\;\;\;\;\;\;
g_{\pi}(z)  =  \frac{2z-1-\mhs/\mts}{1-\mhs/\mts},
\eeq
and  $\zeta$ can be interpreted as $\hnu$, the helicity of the $\nu_{\tn}$ .
The $\rho$-meson can exist in one of two helicity states.
In order not to reduce the sensitivity in the $\tro$ decay  mode, 
it is necessary to spin analyze the $\rho$-meson 
decay products, as discussed in Refs.~\cite{Brau,Rouge}.
Thus $z$ represents a set of three variables:
$\theta^{*}$,  the decay angle of the 
$\rho$-meson in the $\tau$
rest frame; $\psi$,  
the decay angle of the charged $\pi$ in the
$\rho$-meson rest frame;
and  $m_{\rho}$, the mass of the $\rho$-meson.
In the following we use for the $\tro$ decays
the notation introduced in Ref.~\cite{Davier}, 
$d\Gamma \propto( 1 + \zeta P_{\tau} \omega(\theta^{*},\psi,m_{\rho}))d\omega$,
where $\omega(\theta^{*},\psi,m_{\rho})$ is the ratio 
between the polarization-dependent term and the polarization-independent 
term.
  
In the case of the lepton decay modes,  we can describe
the energy spectrum with the Michel parameters~\cite{Michel}
 $\rho$, $\eta$, $\xi$ and $ \delta \xi$.
These parameters were originally conceived to describe $\mu$
decay.  They describe the energy and  angular spectra of the resultant 
lepton with respect to the initial parent spin direction. 
Here the parameters $\rho$ and $\eta$ appear in
the polarization-independent term, while $\xi$ and 
$ \delta \xi $ describe the
polarization-dependent behavior.  
Terms proportional to $\frac{m_{\ell}^2}{E_{\tau}^2}$
may be neglected. 
Then with
$z=\frac{E_{\ell}}{E_{\tau}}$, we obtain the following spectrum for
$\tle$:
\bea
\label{fgleptons}
f_{\ell}(z) &=& 2-6z^2+4z^3 +
                   \rho \frac{4}{9}(-1+9z^2-8z^3)
+\eta\frac{m_{\ell}}{m_{\tau}}(12-24z+12z^2),\nonumber \\
\zeta g_{\ell}(z) &=&    \xi (-\frac{2}{3}+4z-6z^2+\frac{8}{3}z^3)
                 + \delta \xi \frac{4}{9}(1-12z+27z^2-16z^3).
\eea

These decay spectra are combined with the production cross sections to
yield inclusive distributions which can be written as:
\beq
\frac{1}{\sigma}\frac{d^2\sigma(z,\cst,\Pe)}{dz d\cst} \propto
f(z)+\zeta \pt(\cst,\Pe)\cdot g(z).
\eeq
For $\tp$ decays, the sign of
the polarization-dependent term is reversed due to the opposite helicity of
the anti-neutrino. The normalization $\sigma$ in general depends on the decay 
parameters.

The results reported here are based on 
the data
collected by SLD during the period 1993-1995 at a  center of mass energy 
of 91.2~GeV.  The 1993 run accumulated an integrated luminosity of 
1.78 pb$^{-1}$ with an average $\elm$
beam polarization of $(63.0 \pm 1.1)\%$, and the 1994-1995 run 
accumulated 3.66 pb$^{-1}$
with an average polarization of $(77.2 \pm 0.5)\%$.  
A general description of the SLD can
be found elsewhere (~\cite{SLDLIFE}, and references therein).
Charged particle tracking and
momentum analysis are provided by the Central Drift Chamber
(CDC) 
and the CCD-based vertex detector (VXD) 
in an  axial magnetic field of 0.6 T.
Particle energies are measured in the Liquid Argon Calorimeter
(LAC),  which is segmented into projective towers with 
separate electromagnetic (EM) and hadronic sections.
The measured energy and the shape of the energy flow
are used for particle identification.
Additional particle identification is provided
by the \v{C}erenkov Ring-Imaging Detector (CRID), 
 and muons are identified  in
the Warm Iron Calorimeter (WIC).
 
The initial selection of $\tau$-pair candidates
is based on the multiplicity, 
momentum and direction of tracks in the CDC, and on properties of EM
showers in the calorimeter, resulting in a sample of $\Ntpair$ events with
a purity of $98$\% and an efficiency of $80$\% in the fiducial region $|\cst| < 0.74$.
The background contamination and efficiency 
of event selection and decay identification (described below) 
were estimated and parameterized using a Monte Carlo 
(MC) simulation.  
Details on the event selection and simulation
are in Ref.~\cite{SLDLIFE}.
These events are divided into hemispheres by the plane normal to
the event thrust axis, and the hemispheres are treated
independently. Any pair of oppositely charged tracks which is
consistent with originating from a $\gamma$ conversion is removed.
Hemispheres are then required to contain exactly one track, and 
the track is required to have at least one hit in the 
VXD to improve momentum resolution. 

The selection of $\tmu$ in the region  $|\cst| < 0.62$
is performed by associating WIC hits with CDC tracks.
In the region $0.62 < |\cst| < 0.74$, shower information from
the LAC is used instead. A minimum track momentum of 1.6~GeV/c is required.   
This results in a sample of 1143 tracks identified as
muons, with an estimated selection efficiency of $72$\% within the
acceptance, and an estimated purity of 94\%. 
The background comes from $\mu$-pairs (2\%), $2\gamma$ 
events (0.8\%) and mis-identified $\tn$ decays (3.2\%). 
 
For selection of $\tel$  the LAC energy deposition
must be consistent with
that of an electron, or the electron must be identified by the CRID.
The momentum of the track must be greater than 1.6~GeV/c, and
 a quasi-invariant mass calculated using  track momentum and  LAC energy
 clusters is required to be less than 0.5~GeV/c$^2$.
This results in a sample of 948 identified electron
tracks with an estimated efficiency of $63$\% within the 
acceptance and an estimated purity of 96.4\%. 
The background to this mode is composed of Bhabha and 
$2\gamma$ events and mis-identified $\tau$ decays at the levels of
$0.8 \%$, $0.8 \%$ and $2 \%$ respectively.

For the  $\tpiK$ selection, we first reject electron and 
$\mu$ candidates.
Then a candidate is required to have track momentum greater than 3~GeV/c,
and no EM clusters within $10^\circ$ of the thrust axis
that are not associated with a  CDC track.
Here the  calculated quasi-invariant mass  
is required to be less than 
0.3~GeV/c$^2$, and
additional criteria are imposed based on the ratio of LAC energy to track 
momentum.
No attempt is made to separate kaons from pions,
 and a small correction is made in the analysis for the effect of the 
larger kaon mass.
This selection provides a sample of 558 tracks identified as $\tpiK$,
with an efficiency of $58$\% 
within the angular acceptance. 
The purity of the sample is approximately 79\% where the main 
contamination sources are  $\rho$-meson, $e$, and $\mu$ decay
 channels at the rates of $13 \%$, $5 \%$ and $1.4 \%$ respectively.   
The non-tau background is estimated to be
less than 0.5\%.

Events not selected as leptons or pions are candidates for $\tro$ decays.
Hemispheres are then
categorized according to the
number of EM  clusters found within $20^\circ$ of the event
thrust axis, and according to
whether such clusters are associated or unassociated with
the charged track in the corresponding hemisphere.  
The charged track momentum and
the EM cluster energies are used to calculate a $\rho$-meson energy and its 
invariant mass,
and the mass is required to fall in the range between 0.44 and 1.2
GeV/$c^2$.
This selection results in a sample of 1295 identified $\rho$-meson decays, 
with an efficiency of $60$\% 
over the fiducial region. The purity of the sample is
approximately 76\%, 
where background is dominated  by decays to  
$\pi 2\pi^{0}$  ($14.7 \%$), K$^*$ (2.4\%),
and single $\pi$ or K (2.4\%).
The non-tau background is estimated to be less than 0.3\%. 

In Fig.~2(a), 
the energy spectrum of $\tpiK$ decays is plotted
separately for different combinations of the production angle and the sign of
$\Pe$.
\begin{figure}[ptb]
\centerline{\epsfig{file=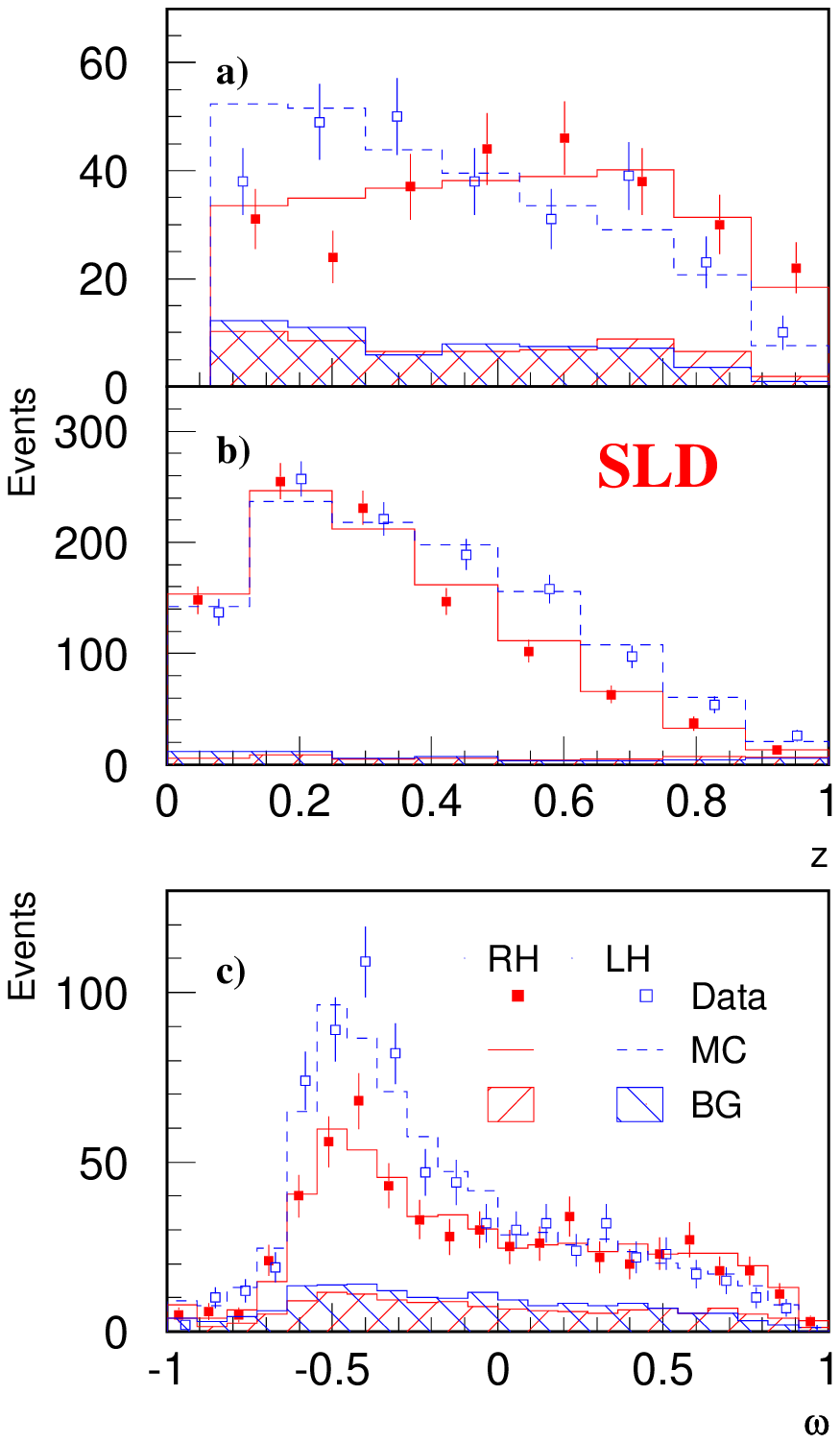
}}
\caption{(a) $\tpiK$, (b) $\tle$ and (c) $\tro$ decay  spectra.
The solid red 
squares (line) represent the sum of the measured  (MC) 
spectra for $\tau$ decays in the forward direction with 
$\Pe < 0$ and in the backward direction with $\Pe > 0$, and the
open blue 
squares (dashed line) are the measured (MC) 
sum of the spectra for $\tau$ decays in the backward direction
with $\Pe < 0$ and in the forward direction with $\Pe > 0$.
The hatched regions represent the estimated backgrounds in the 
two combinations. The MC was generated with the SM values for the $\tau$ decay parameters.}
\label{pispect}
\end{figure}
For left-handed incident $\elm$ and $\tm$ emitted at forward 
polar angles, or for right-handed beam and $\tm$ 
in the backward region, the $\tm$ are predominantly left-handed 
and the $\pi$ energy spectrum is expected to be relatively soft.
For the two opposite combinations of $\Pe$ and $\cst$,
the spectrum should be harder since the decaying $\tm$ 
are predominantly right-handed. Fig.~2(c) 
shows the same comparison for the $\tro$ decays. 
The difference is expected to be less obvious for
the three-body decays $\tle$, but is still quite visible  in
Fig.~2(b). These plots also demonstrate the good agreement between
data and MC, as well as the low and helicity-symmetric backgrounds. 
 
The $\nut$ helicity 
and the Michel parameters $\rho$, $\eta$, $\xi$ 
and $\delta\xi$ are determined using a maximum likelihood fit to
the $\tau$ production angle and the energy  spectra of the decay
channels $\tpiK$ and $\tle$ ($\omega$ spectrum for $\tro$).
The following expression is minimized: 
\beq
\label{ML}
W=-2\sum_{decays} \ln \left\{ \frac{1}{\spr}
\frac{d^2\spr}{d\cst\:dz} \right\}.
\eeq
The sum in Eq.~\ref{ML} runs  over all 
 $\tel$, $\tmu$, $\tro$ or $\tpiK$ candidates. 
The fit function includes effects of $\gamma$ exchange and $\gamma-\Z$ 
interference, 
radiation, detector resolution, efficiency and backgrounds. 
The dependence of MC efficiencies and backgrounds on
$z$ ($\omega$) and $\cst$ 
is parameterized using low order polynomials.  
The effect of initial and final state radiation is determined from
the ratio of the spectrum 
for  MC events generated using KORALZ~4.0~\cite{KORAL} (including radiation) 
to the spectrum 
for the Born level cross-sections.  
Effects of detector resolution are studied using MC for each input
variable (i.e. track momentum or $\omega$, and $\tau$  direction).
Fits to
multiple Gaussians are performed to model both the core and tails of the
resolution distributions, and  these functions are convoluted  with the
theoretical expression.
Since the spectra are different for decays
of left- and right-handed taus, all the correction functions are divided into
four categories corresponding to combinations of positive or negative $\elm$
beam polarization and the forward or backward half of the detector.  
In all cases the analytic functions are required to be good 
representations of the MC distributions.

 These measurements are dominated by statistical errors.
Table~I summarizes the systematic errors. 
\begin{center}
{\footnotesize TABLE I. Systematic uncertainties (in units of $10^{-2}$).}
\begin{tabular}{lccccccccc} \hline\hline
 & $h_\nu^{\pi}$ & $ h_\nu^{\rho}$  & 
 $\rho^e$ &  $\rho^{\mu}$ &  $\eta^{\mu}$ &
 $\xi^e$ &  $\xi^{\mu}$ &               
 $\delta\xi^e$ &  $\delta\xi^{\mu}$ \\ \hline
 Selection                     & 0.7 & 2.7 & 2.7 & 4.5  & 13.8  & 2.6 & 4.3  & 4.4 & 2.0 \\
 Background                    & 1.1 & 1.3 & 1.2 & 13.3 & 41.7  & 1.1 & 13.1 & 0.7 & 6.3 \\
 K fraction
                               & 0.8 &     &     &      &      &     &      &     &       \\ 
 Rad. corr.                    & 0.5 & 0.4 & 0.4 & 0.4  & 0.8  & 0.2 & 0.1  & 0.2 & 0.1 \\
 Resolution 
                               & 1.3 & 1.5 & 3.8 & 2.6  & 5.1  & 4.9 & 2.2  & 5.9 & 2.9 \\
 Beam $\Pe$                    & 0.9 & 0.6 & 1.6 & 0.4  & 4.5  & 0.9 & 2.1 & 0.7  & 0.4 \\
 $A_e$, $A_\tau$~\cite{Al}     & 0.7 & 0.8 & 0.3 & 0.2  & 1.4  & 1.6 & 2.4  & 1.6 & 0.8   \\ 
 TOTAL                         & 2.4 & 3.5 & 5.1 & 14.3 & 44.5 & 5.9 & 14.4 & 7.6 & 7.2 \\ \hline 
\hline\hline
\end{tabular}
\label{systab}
\end{center}
To investigate these, the parameterizations described above are modified by
their uncertainties as determined from the MC,  the fit
is redone to obtain new
values of $\hnu$, $\rho$, $\eta$, $\xi$ and $\delta\xi$, and changes in the 
fitted values are taken as the systematic errors.  
The validity of these errors has been checked by comparing data and MC 
distributions, for example,  of cluster energies, number of associated and 
unassociated clusters, and $\pi^0$ and $\rho$-meson reconstructed masses. 
The observed number of events and the calculated efficiencies
and backgrounds are consistent with measured branching ratios~\cite{PDG}.
The uncertainty in the fraction of kaons in the $\tpiK$ sample 
affects the correction for the kaon mass. 
The errors on radiative corrections are dominated by MC statistics.
The beam polarization uncertainties are as quoted above. 
Each error listed in Table~I
may include  several related contributions.
These errors are combined taking into account any correlation between 
the input parameters.

The measured values for $\hnu$ and the  Michel parameters $\rho$, $\eta$, 
$\xi$ and $\delta\xi$  are given in 
Table ~II. 
\begin{center}
{\footnotesize 
TABLE II. The measured $\hnu$ and  Michel parameters including statistical\\
and systematic errors, given  by decay channel and as combined results, \\
compared with the Standard Model (SM) predictions.}
\begin{tabular}{lcccc} \hline \hline
    & SM & $\tpiK$ & $\tro$ & hadrons combined   \\ \hline
$\hnu$ & \hspace{-3.5mm}$-1$ & \hspace{-3.5mm}$\PIHNU \pm \PIHNUST  \pm \PIHNUSY$
       & \hspace{-3.5mm}$\ROHNU \pm \ROHNUST  \pm \ROHNUSY$
       & \hspace{-3.5mm}$\HNU   \pm \HNUST  \pm \HNUSY$  \\ \hline \hline
     & & $\tel$ & $\tmu$ & $\tle$ combined   \\ \hline
$\rho$ & $\frac{3}{4}$ & $\ERHO \pm \ERHOST  \pm \ERHOSYS$ 
       & $\MRHO \pm \MRHOST  \pm \MRHOSYS$
       & $\NURHO \pm \NURHOST    \pm \NURHOSYS$ \\
$\eta$ & 0 &  
       & \hspace{-3.5mm}$\META \pm \METAST  \pm \METASYS$
       &  \\
$\xi$  & 1 & $\EXI \pm \EXIST  \pm  \EXISY$
       & $\MXI \pm \MXIST  \pm  \MXISY$
       &  $\NUXI \pm \NUXIST  \pm  \NUXISYS$\\
$\delta\xi$ & $\frac{3}{4}$ & $ \EXIDELTA \pm \EXIDELTAST  \pm \EXIDELTASY$
            & $\MXIDELTA \pm \MXIDELTAST  \pm \MXIDELTASY$
            & $\NUDX \pm \NUDXST  \pm \NUDXSYS$ \\ \hline \hline
\end{tabular}
\label{leptons-res}
\end{center}
For the combined $\tle$ fit the correlation coefficients are 
$C_{\rho-\xi}=0.06$, $C_{\rho-\delta\xi}=0.03$ and
$C_{\xi-\delta\xi}=0.08$,  
while for $\tmu$ the $\eta^{\mu}$ and $\rho^{\mu}$ parameters are highly 
correlated (0.92).
Note that $\eta$ could not be measured separately
for $\tel$ due to the small $m_e/m_{\tau}$ term.
One could improve the $\eta$ measurement by
including the $\tel$ data and assuming universal couplings for $e$ and
$\mu$ in the charged current.
In this case we get:
$\rho=\RHO \pm \RHOST \pm \RHOSYS$,
$\eta=\ETA \pm \ETAST \pm \ETASYS$,
$\xi=\XI \pm \XIST \pm \XISYS$,
$\delta\xi=\XIDELTA \pm \XIDELTAST \pm \XIDELTASY$.
However this assumption 
is inconsistent because generally a non-zero value for $\eta$
would imply $\rho$ to be  non-universal~\cite{Fetscher}.  

The results are consistent with the  $V-A$ predictions and are in good 
agreement with other experimental 
results [5-10]. 
This is the first measurement using beam polarization and we have
measured not only the magnitude but also the sign of all
pseudoscalar and parity violating quantities.
Because of the unique method, 
significant new measurements have been derived from a relatively 
small number of events. 
One can also interpret the results from 
leptonic decay channels~\cite{MOURAD} as
${\mbox h^{leptons}_{\nu_{\tau}}=\LHNU \pm \LHNUST \pm \LHNUSY}$,
and combined with  the hadronic decay channels one obtains 
${\mbox h^{lept+had}_{\nu_{\tau}}=\LHHNU \pm \LHHNUST \pm \LHHNUSY}$.

We thank the personnel of the SLAC accelerator department and the
technical staffs of our collaborating institutions for their
outstanding efforts on our behalf.
This work was supported by the U.S. Department of Energy 
and National Science Foundation, 
the UK Particle Physics and Astronomy Research Council,
the Istituto Nazionale di Fisica Nucleare of Italy,
the Japan-US Cooperative Research Project on High Energy Physics,
and the Korea Science and Engineering Foundation. 


\begin{thebibliography}{99}
\bibitem{MOURAD} The leptonic charged weak interactions for $e$ and $\mu$ have
been measured with sufficient 
precision to exclude interactions
different from a pure $V-A$ at  the $\ell\nulb$ vertex~\cite{Fetscher,PDG}. 
Therefore, one can assume
 $V-A$ at the $\ell\nulb$ vertex
and allow a combination of $V-A$ and $V+A$ at the $\nut\t$ 
vertex~\cite{Privitera}. 
This assumption reduces the
number of free parameters 
to one as in the hadronic decay channels.
In this case the relations between $\hnu$ and $\rho, \xi,  \delta \xi$ are the 
following:
$\hnu=1-\frac{8}{3}\rho=\xi-2=1-\frac{8}{3}\delta\xi$; $\eta=0$.
\vspace{-0.2cm}
\end{thebibliography}
\end{document}